\documentclass[prd, twocolumn, eqsecnum, floatfix, letterpaper, superscriptaddress, nofootinbib]{revtex4}
\usepackage[utf8]{inputenc}
\usepackage{color, ulem, amsmath, amssymb, graphicx, bm, times, microtype, csquotes, booktabs, natbib}
\usepackage[varg]{txfonts}
\usepackage[colorlinks, pdfborder={0 0 0}]{hyperref}
\usepackage{subcaption}
\usepackage{ragged2e}
\usepackage{algorithm}
\usepackage{algorithmic}
\usepackage{tabularx}
\usepackage{float}
\newcommand{\e}{\mathrm{e}}

\newcommand{\red}{\color{black}}

\begin{document}
\title{Testing compact binary nature in eccentric orbits with third-generation detectors}
\date{\today} 

\newcommand{\IITM}{Department of Physics, Indian Institute of Technology Madras, Chennai 600036, India}
\author{Syed U. Naqvi}
\email{syed.naqvi@ifj.edu.pl}
\affiliation{Institute of Nuclear Physics, Polish Academy of Sciences, Radzikowskiego 152, 31-342 Kraków, Poland}
\affiliation{\IITM}
\author{Kaushik Paul} 
\affiliation{Nikhef, Science Park 105, 1098 XG Amsterdam, The Netherlands}
\author{Chandra Kant Mishra} 
\affiliation{\IITM}

\begin{abstract}
The spin-induced quadrupole moment (SIQM) of a compact object provides a direct null test of the Kerr hypothesis. Most existing gravitational-wave tests of SIQM assume quasi-circular binaries, although dynamical formation channels can produce measurable eccentricity that may persist in the sensitive band of third-generation (3G) detectors. We extend the symmetric SIQM test to eccentric, aligned-spin binaries using a post-Newtonian spinning-eccentric phasing model, and forecast its measurability with Cosmic Explorer (CE) and Einstein Telescope (ET). Single-event Fisher-matrix forecasts show that the impact of eccentricity on SIQM measurement precision depends on the binary parameters and detector configuration, rather than uniformly degrading the measurement. A population study with a truncated eccentric expansion indicates that ET's additional low-frequency eccentric cycles tighten $\Delta\kappa_s$ by roughly a factor of two relative to CE (median $0.11$--$0.12$ versus $0.20$--$0.23$); the absolute values are indicative only. We further show that quasi-circular template banks can lose significant effectualness for moderately eccentric signals, motivating the development of eccentric-spin template families. Accounting for eccentricity is therefore important both for SIQM measurements and for avoiding systematic losses in searches, and it strengthens tests of the nature of compact objects with 3G gravitational-wave detectors.
\end{abstract}\label{abstract}

\maketitle

\section{Introduction}\label{intro}

Gravitational-wave (GW) observations of compact-binary coalescences provide a unique observational route to testing the nature of black holes (BHs) and other compact objects in the strong-field, dynamical regime \cite{LIGOScientific:2016aoc,LIGOScientific:2021djp,YunesSiemens:2013,Berti:2015}.  In general relativity, an isolated astrophysical BH is described by the Kerr solution and is completely characterized by two numbers its mass and spin \cite{Israel:1967,Carter:1971,Hawking:1972}.  This implies a highly restrictive multipolar structure in the sense that mass and current multipole moments are fixed functions of the two Kerr parameters \cite{Geroch:1970,Hansen:1974}.  Any statistically significant departure from this structure, that cannot be attributed to waveform systematics or environmental effects, may hint toward either non-Kerr compact objects or new physics beyond the standard binary-BH description \cite{Ryan:1995}.

Following this, the current detections have motivated a broad program of compact-object null tests with GWs to ascertain the nature of compact objects.  Several classes of exotic compact objects (ECOs), including boson stars, gravastars, fuzzballs, wormholes, and other horizonless or near-horizon alternatives, can be sufficiently compact to mimic BHs in parts of the inspiral, merger, or ringdown signal \cite{Cardoso:2017cqb,Cardoso:2019rvt}.  However, their internal structure and boundary conditions can differ from those of Kerr BHs, leading to potentially observable deviations in tidal deformability, horizon absorption, quasinormal-mode spectra, or spin-induced multipole moments \cite{Cardoso:2017cqb,Cardoso:2019rvt,Maggio:2021ans}.  In this context, spin-induced multipole moments are particularly useful because they enter the inspiral dynamics and can be constrained before the highly nonlinear merger regime.

The leading spin-induced multipole relevant for compact-binary tests is the quadrupole moment.  For the $A$th binary component, we write
\begin{equation}
Q_A=-\kappa_A \chi_A^2 m_A^3,
\end{equation}
where $m_A$ is the mass, $\chi_A$ is the dimensionless spin, and $\kappa_A$ is the spin-induced quadrupole coefficient.  For Kerr BHs, $\kappa_A=1$.  For neutron stars, $\kappa_A$ depends on the equation of state and typically exceeds the Kerr value, while for several ECO models the deviation can be significantly larger \cite{Laarakkers:1997hb,Poisson:1997ha,Krishnendu:2017shb,Krishnendu:2018nqa}.  Boson stars provide an important example: their quadrupole moments and tidal deformabilities depend on the bosonic-field model and self-interaction parameters, and can differ strongly from the Kerr expectation \cite{Ryan:1996nk,Adam:2022pdb}.  Gravastars and other horizonless compact objects similarly provide test beds in which the absence of a classical event horizon and the presence of matter or effective surfaces can modify the multipolar response relative to BHs \cite{Cardoso:2017cqb,Cardoso:2019rvt}.

GW measurements of $\kappa$ therefore provide a physically interpretable null test of the binary-BH hypothesis. The first applications of this idea showed that spin-induced quadrupole terms in the inspiral waveform can be used to constrain deviations from the Kerr value and to distinguish BH binaries from systems containing neutron stars or ECOs \cite{Krishnendu:2017shb,Krishnendu:2018nqa}. Subsequent Bayesian analyses, incorporating effects such as double spin-precession and higher harmonics, have demonstrated that these measurements are robust against waveform systematics and can be applied to actual GW events \cite{Divyajyoti:2024siqm}. Going beyond the quadrupole, the spin-induced octupole moment, which enters the waveform at higher post-Newtonian order, has also been implemented for testing the Kerr nature. A Bayesian framework for measuring the octupole moment was developed in Ref.~\cite{Saini:2024octupole}, which applied the test to events from the third GW catalog and obtained full Bayesian constraints on the octupole deviation parameter. More recently, the analysis of GW241011 enabled a simultaneous test of both quadrupole and octupole deviations from the Kerr prediction, placing the first observational constraints on the spin-induced octupole moment of a compact binary \cite{Das:2026octupole}. Looking ahead, third-generation detectors such as Cosmic Explorer and Einstein Telescope will be able to measure the symmetric combination of octupole coefficients to within $\lesssim 2$ \cite{Saini:2024octupole}. Hence, SIQM (and higher-multipole) measurements can go beyond phenomenological consistency checks and, with future detections, be strong candidates for mapping onto concrete regions of ECO parameter space.

In practice, the two individual coefficients $\kappa_1$ and $\kappa_2$ are difficult to measure simultaneously because of correlations with masses, spins, and other intrinsic parameters.  It is therefore useful to introduce the symmetric and antisymmetric combinations:
\begin{equation}
\kappa_s=\frac{\kappa_1+\kappa_2}{2}, \qquad
\kappa_a=\frac{\kappa_1-\kappa_2}{2}.
\end{equation}
For a binary composed of two Kerr BHs, $\kappa_s=1$ and $\kappa_a=0$.  A common null-test implementation measures $\kappa_s$ while fixing $\kappa_a=0$, corresponding to the symmetric SIQM test.  Equivalently, one may introduce deviation parameters $\delta\kappa_A$ through $\kappa_A=1+\delta\kappa_A$, so that the Kerr limit is recovered when $\delta\kappa_A=0$ \cite{Krishnendu:2017shb,Krishnendu:2018nqa,LIGOScientific:2021djp}.  In this work, we follow the symmetric-test strategy and focus on the measurability of $\kappa_s$.

Most existing SIQM-based tests have been developed for quasi-circular compact binaries.  This is a natural starting point because GW radiation tends to circularize binaries formed through isolated stellar evolution before they enter the frequency band of ground-based detectors.  However, eccentricity is expected in several dynamical formation channels, including dense stellar clusters, galactic nuclei, hierarchical triples, and active galactic nucleus disks \cite{Samsing:2018isx,Rodriguez:2018pss,Antonini:2017tct,Tagawa:2020qhl}.  Eccentricity is therefore a key tracer of binary formation history and has already been used in Bayesian analyses of observed GW events.  Early studies of GWTC-1 found the detected binary black holes to be consistent with zero eccentricity while placing 90\% upper limits of order $e_{10\,{\rm Hz}}\lesssim 0.02$--$0.05$ \cite{Romero-Shaw:2019bph}.  Subsequent analyses reported evidence or support for eccentric interpretations of selected events, including GW190521 and GW200208\_222617, and explored the possibility that multiple LVK events may carry eccentric signatures \cite{Romero-Shaw:2020thy,Romero-Shaw:2022zhr,Romero-Shaw:2025abc}.  More recently, systematic reanalyses of the growing GW catalog and dedicated event-specific studies have extended this picture to additional binary-black-hole events and to neutron-star-containing binaries \cite{Planas:2025ecc,Kacanja:2025ecc,Jan:2026ecc}.  These studies show that eccentricity is not only a theoretical formation-channel indicator, but also a measurable parameter in current and future GW observations.  However, the presence of eccentricity introduces additional phase modulations that can mimic or mask spin-induced effects, potentially biasing SIQM constraints if eccentricity is neglected.  A quantitative understanding of how eccentricity affects the measurability of spin-induced multipole moments is therefore essential for robust tests of the Kerr hypothesis in dynamically assembled binaries.

The importance of eccentricity is further strengthened by the possibility that it may encode physics beyond standard stellar dynamics.  For example, ultralight-boson clouds around spinning BHs can exchange energy and angular momentum with a binary companion through resonant transitions, modifying the orbital evolution and potentially producing measurable eccentricities in low-frequency bands \cite{Boskovic:2024qtr}.  Such examples show that eccentricity can carry information about the astrophysical environment, formation channel, or new fields coupled to compact objects \cite{Loutrel:2020ecc,Gupte:2025ecc}.  It is therefore important to understand how eccentricity interacts with SIQM measurements, and whether it improves the distinguishability of Kerr BHs from neutron stars, boson stars, gravastars, or other ECOs, or instead introduces additional parameter degeneracies.  Ignoring eccentricity in waveform models can bias parameter estimation and even mimic spin-induced effects, potentially compromising null tests of the Kerr hypothesis \cite{Divyajyoti:2024blind,Bhat:2024eccGR}.

Recently, Krishnendu~et~al.~\cite{Krishnendu:2026eccSIQM} investigated the impact of neglected eccentricity on SIQM measurements using the TaylorF2Ecc waveform model, where the spin-induced quadrupole coefficient $\kappa_s$ enters only through the circular-spin terms and eccentricity corrections are treated additively. Their work quantified the systematic biases in $\kappa_s$ recovery for selected source parameters when circular templates are used for eccentric signals. In this paper, we employ the more comprehensive waveform model of Sridhar~et~al.~\cite{Omkar2025}, which provides phasing expressions accurate up to 3PN order in spin and \(\mathcal{O}(e_0^8)\) in eccentricity, and incorporates \(\kappa_s\) in both the circular and eccentric parts of the phasing, thereby capturing the full coupling between spin-induced deformations and orbital eccentricity. Using Fisher-matrix forecasts, we extend the analysis to the full \(\chi_1\)--\(\chi_2\) spin plane for a range of masses, eccentricities, and third-generation detector configurations. This allows us to assess how eccentricity modifies the intrinsic measurability of SIQM parameters and whether it ultimately strengthens or compromises the SIQM test for third-generation detectors.

\subsection{Summary of the current work}\label{subsec:this_work}

In this paper, we extend the spin-induced quadrupole-moment test of compact-binary nature to eccentric inspirals utilizing the latest spinning eccentric waveform.  We employ the aligned-spin eccentric post-Newtonian phasing model of Sridhar~et~al.~\cite{Omkar2025}, which provides closed-form time- and frequency-domain phasing expressions including spin effects up to third post-Newtonian order and eccentric corrections expanded in powers of the initial eccentricity.  We use this existing eccentric spinning phasing framework to forecast how well the SIQM parameter $\kappa_s$ can be measured for eccentric compact binaries.  Throughout the analysis, we focus on the symmetric SIQM parameter $\kappa_s$ and set $\kappa_a=0$, corresponding to the standard symmetric null-test configuration.

We carry out three complementary studies.  \textit{First}, we compute single-event Fisher-matrix forecasts for the uncertainty $\Delta\kappa_s$ and study its dependence on total mass, component spins, and initial eccentricity.  This analysis quantifies how eccentricity changes the measurability of the spin-induced quadrupole coefficient, and how the errors on $\kappa_s$ and $e_0$ depend on spin alignment.  \textit{Second}, we perform a population-level study and compare the projected sensitivity of CE and ET for measuring $\kappa_s$ across a population of eccentric compact binaries.  This allows us to assess how detector sensitivity and source population assumptions affect the expected constraints on the binary-BH nature. \textit{ Third}, in Sec.~\ref{sec:temp_bank} we estimate the importance of inclusion of eccentricity in SIQM search scenarios by computing fitting factors $(\mathcal{FF})$ between eccentric aligned-spin signals and an aligned-spin quasi-circular template bank for two different values of $\kappa_{1,2}=1$ (the Kerr BH case) and 10. The top panel of Figure~\ref{fig:ff_and_frac_recovered} shows the variation of $\mathcal{FF}$ as a function of initial eccentricity ($e_{10}$ for aLIGO and $e_5$ for CE and ET), while the bottom panel quantifies the fraction of eccentric injections that has been recovered by the quasi-circular template bank with a $\mathcal{FF}>0.97$.

Because observational bounds on $\kappa_s$ can be translated into constraints on rotating ECO models \cite{Krishnendu:2025rud}, our forecasts also indicate how eccentricity and third-generation detector sensitivity may improve the model-discriminating power of SIQM-based compact-object tests.  In particular, improved measurements of $\kappa_s$ can help determine whether future eccentric compact-binary observations are consistent with Kerr BHs or whether they require interpretations involving neutron stars, boson stars, gravastars, or other black-hole mimickers.

The paper is organized as follows.  In Sec.~\ref{sec:waveformmodel}, we summarize the waveform model, the spin-induced quadrupole parametrization, and the eccentric aligned-spin phasing adopted in this work.  In Sec.~\ref{sec:PE}, we describe the Fisher-matrix formalism and the detector configurations.  In Sec.~\ref{sec:single_event_results}, we present single-event forecasts for $\Delta\kappa_s$ and discuss its dependence on masses, spins, and eccentricity.  In Sec.~\ref{sec:population_synthesis}, we present the population-level CE--ET comparison.  In Sec.~\ref{sec:temp_bank}, we discuss the $\mathcal{FF}$ computation of eccentric aligned-spin injections and quasi-circular aligned-spin template bank generated using the inspiral-only, frequency domain waveform presented in Sec. III C of Ref.~\citep{Omkar2025}.  Finally, in Sec.~\ref{sec:conclusion}, we summarize our conclusions and discuss future extensions, including precession, larger eccentricities, and full Bayesian inference.

\section{Post-Newtonian Modeling of Spin-Induced Multipole Moments}\label{sec:waveformmodel}

The inspiral phase of a compact binary system is accurately modeled by the post-Newtonian (PN) formalism \cite{Blanchet:2013haa}. PN waveforms have been used for detection and parameter estimation from compact binaries with nonspinning constituents in quasi-circular orbits since the early 2000s \cite{Cutler:1994ys,Buonanno:2009zt}. Many improvements have since incorporated higher-order spin effects, eccentricity, and precession \cite{Kidder:1995zr,Apostolatos:1994mx,Moore:2016qxz}. We focus on a frequency-domain waveform where the spins are (anti-)aligned with respect to the orbital angular momentum for a binary on an eccentric orbit \cite{Omkar2025}. We use a state-of-the-art frequency-domain eccentric waveform for compact binaries with (anti-)aligned spin components, having spin-orbit effects in the phasing up to 4PN (leading effect at 1.5PN), spin-spin effects up to 3PN (starting at 2PN), and the leading cubic-spin terms at 3.5PN. The amplitude includes spin effects up to 2PN.

The baseline waveform used for our analysis consists of a 2PN circular amplitude with spin corrections and a 4PN circular phase with spinning corrections. To this, we add:
i) Non-spinning 3PN eccentric terms (expanded in $e_0^2$), and
ii) Spinning 3PN eccentric terms (expanded in $e_0^2$, $e_0^4$, $e_0^6$, and $e_0^8$).

We have used a leading (second) harmonic (quadrupolar mode) waveform and its PN corrections in the amplitude augmented by the eccentric correction (in orders of $e_{0}^{2}$). The presence of higher modes is neglected and schematically the waveform reads as
\begin{equation} \tilde{h}(f)=\frac{M^2}{D_L} \sqrt{\frac{5\,\pi
\,\eta}{48}}\sum_{n=0}^{4} V_2^{n-7/2}\,C_{2}^{(n)}\,\mathrm{e}^{\mathrm{i}
\,\big({2\,\Psi}_\mathrm{SPA}(f/2)-\pi/4\big)}\, . \label{eq:waveform1}
\end{equation}

Here \(M\), \(\eta\), and \(D_{L}\) denote the total mass, symmetric mass-ratio, and luminosity distance of the binary, respectively.  The coefficients \(C_{2}^{(n)}\) represent the amplitude corrections to the quadrupolar harmonic at \((n/2)\) PN order.  The factor \(V_{2} = (\pi M f)^{1/3}\) encodes the dependence on the gravitational-wave frequency \(f\). The waveform phase \(\Psi_{\rm SPA}(f)\) is decomposed as
\[
\Psi_{\rm SPA}(f) \;=\; \Psi_{\rm circ}^{\rm 4PN}(f) \;+\; \Psi_{\rm ecc}(f)\,,
\]
where \(\Psi_{\rm circ}^{\rm 4PN}(f)\) is the spinning circular component accurate to 4 PN order \cite{Bohe:2013cla,Bohe:2015ana,Marsat:2014xea}, and \(\Psi_{\rm ecc}(f)\) contains the eccentric corrections up to 3 PN.  In particular:

\begin{itemize}
  \item \(\Psi_{\rm ecc}^{\rm NS}(f)\) (non-spinning eccentric) is taken from \cite{Moore:2016qxz}, expanded through \(\mathcal{O}(e_{0}^{2})\).
  \item \(\Psi_{\rm ecc}^{\rm S}(f)\) (spinning eccentric) follows \cite{Omkar2025}, expanded through \(\mathcal{O}(e_{0}^{8})\).
\end{itemize}

Thus, our analysis employs two phase models:
\begin{align*}
\text{i)}\quad \Psi_{\rm SPA}(f) &= \Psi_{\rm circ}^{\rm 4PN}(f) + \Psi_{\rm ecc}^{\rm NS}(f)\,, \\
\text{ii)}\quad \Psi_{\rm SPA}(f) &= \Psi_{\rm circ}^{\rm 4PN}(f) + \Psi_{\rm ecc}^{\rm S}(f)\,.
\end{align*}
where ``NS'' denotes non-spinning eccentric corrections and ``S'' denotes spinning eccentric corrections. Note that in $(i)$ $\kappa_{s}$ enters only in the circular part and in $(ii)$ it enters in both circular and eccentric part. The functions \(C^{(n)}_{2}\) and the phasing---each explicitly depending on the spin-induced quadrupole parameters (\(\kappa_s\), \(\kappa_a\)) and octupole parameters (\(\lambda_s\), \(\lambda_a\)) at their respective PN orders---are provided in the supplemental material of \cite{Omkar2025}.

Poisson \cite{Poisson:1997ha} evaluated the effect of the leading spin-induced multipole moment (mass-type quadrupole, \(M_2 = -M^3 \chi^2\)) on the phasing of gravitational waves from binary black hole systems at 2PN order, where \(M\) and \(\chi\) denote the mass and dimensionless spin parameter of each binary component, and the negative sign indicates that spin induces oblateness in the black hole. The 3PN corrections to this term were computed in \cite{Bohe:2015ana}. At 3.5PN order, the subleading spin-induced multipole moment (current-type octupole, \(S_3 = -M^4 \chi^3\)) begins to contribute, as computed in \cite{Marsat:2014xea}. Note that at 2PN (3.5PN), the spin dependencies of the spin-induced multipole moments \(M_2\) (\(S_3\)) arise, corresponding to the orders where quadratic-in-spin (cubic-in-spin) terms first appear in the gravitational-wave phase. Similarly, in the spinning-eccentric expansion of Sridhar et al.\ \cite{Omkar2025}, the spin-induced multipole moments also appear at the same respective PN orders. The interplay between eccentricity and spin-induced multipole terms affects parameter estimation in such systems.

The \(M_2\) and \(S_3\) relations described before can be generalized from black hole binary constituents to non-BH compact objects with the introduction of the coefficients that characterize the degree of deformations. These relations can be rewritten as \(M_{2} = -\kappa M^3 \chi^2\) and \(S_3 = -\lambda M^4 \chi^3\), where the coefficients \(\kappa\) and \(\lambda\) have a unity value for Kerr black holes (guided by the `no-hair theorem'). But for exotic objects different from black holes, the value of \(\kappa\) and \(\lambda\) is no longer unity. In \cite{Laarakkers:1997hb,Krishnendu:2018nqa}, it was shown that owing to the neutron star's equation of state and mass, the values range between \(\sim 2\text{--}14\) and \(\sim 4\text{--}30\), respectively. In other cases of spinning boson stars, the values of spin-induced multipole moments \(\kappa\) (\(\lambda\)) can take values between \(\sim 10\text{--}150\) (\(\sim 10\text{--}200\)) as shown in \cite{Ryan:1996nk,Adam:2022pdb}. For gravastars, in the thin-shell model, the values of spin-induced multipole moments have been evaluated in \cite{Uchikata:2016qku}. Interestingly, if the observed values of spin-induced moments are different from black hole values, it can potentially hint toward exotic compact objects. Also, if their posterior distributions of the observed value are found to be peaking at 1 with a width, the corresponding errors can be used to set an upper bound on the allowed value of the parameter pertaining to a particular system. Earlier works have computed spin-induced moments for binaries in circular orbits; we extend those to eccentric orbits using the semi-analytical estimation technique of the Fisher matrix, with the analysis being detailed in the next section.

\section{Parameter Estimation Framework}\label{sec:PE}

Given an accurate model for the signal of interest and the expected detector sensitivity, the Fisher matrix formalism can be used to compute $1\sigma$ uncertainties on the signal's parameters \cite{Finn:1992wt,Cutler:1994ys}. This approach relies on the assumptions of a high signal-to-noise ratio (SNR) and Gaussian-stationary detector noise. While these assumptions significantly limit the applicability and robustness of the Fisher matrix for realistic data analysis (especially for low-SNR signals or non-Gaussian noise), it remains a valuable tool for prospective uncertainty forecasting \cite{Vallisneri:2007ev}. We employ this approach to estimate uncertainties on parameters characterizing eccentric binary systems and to evaluate the influence of spin-induced multipole moments on parameter estimation accuracy in compact binaries. For completeness, we provide a brief overview of the Fisher matrix formalism below; more details can be found in \cite{Vallisneri:2007ev}.

A gravitational wave detector data is composed of the signal, $h(t;\theta_i)$ (here $\theta_{i}$ represent the set of parameters characterizing the signal) and the background noise, $n(t)$ written as:
\begin{eqnarray} {s}(t)=h(t;{\theta_i})+n(t), \label{eqn3} \end{eqnarray}

The goal of parameter estimation is to use waveforms coming from the theory and provide uncertainty bounds on the parameters \(\theta_{i}\). As there is noise in the detector, the measured parameters \(\theta_{i}\) fluctuate about the true value, leading to errors associated with their measurement. Therefore, the measured value is \(\theta_i = \theta_i^{\rm true} \pm \Delta \theta_i\), where \(\theta_i^{\rm true}\) denotes the true value of the parameter and \(\Delta \theta_i\) represents the error associated with the measurement due to noise. Once a measurement is done, we are interested in the probability distribution function for \(\theta_i\) given the signal \(s(t)\), \(p(\theta_i \mid s)\). For Gaussian noise and under the limit of high signal-to-noise ratios, the posterior probability is related to the errors \(\Delta \theta_{i}\) as
\begin{eqnarray} p({\theta_i}|s) \propto
\e^{-\frac{1}{2}(\Gamma_{jk}\Delta{\theta_{j}} \Delta{\theta_{k}})}, \label{eqn5}
\end{eqnarray}
where $\Gamma_{ij}$ is the Fisher information matrix \cite{Finn:1992wt,Cutler:1994ys} defined as follows:
\begin{eqnarray} \Gamma_{ij}=2 \int^{f_{\rm upper}}_{f_{\rm lower}} df
\frac{{\tilde h_{i}}(f){\tilde h_{j}}^{\ast}(f)+ {\tilde h_{j}}(f){\tilde
h_{i}}^{\ast}(f)}{S_{n}(f)}, \label{eqn6} \end{eqnarray}
where $S_{n}(f)$ is the noise power spectral density (PSD) of the detector and the quantity $\tilde{h}_{i}\equiv{\partial\tilde{h}(f;{\theta_{i})}}/{\partial {\theta_i}}$ is the derivative of the waveform evaluated at the true value of the parameter ${\theta}_{i}={\theta}_{i}^{\rm true}$. The covariance matrix ($\Sigma_{ij}$) and the associated $1\sigma$ error ($\sigma_{i}\equiv\Delta\theta_i$) on each parameter $\theta_{i}$ are inferred from the inverse of the Fisher information matrix and the square root of the diagonal entries of the covariance matrix, respectively. Namely,
\begin{eqnarray} \sigma_{i}=\sqrt{\Sigma_{ii}}. \label{eqn7} \end{eqnarray}
The Fisher matrix is computed over the eight-dimensional parameter vector
\begin{equation}
\bm{\theta}=\{t_c,\ \phi_c,\ \mathcal{M}_c,\ \eta,\ \chi_1,\ \chi_2,\ \kappa_s,\ e_0\},
\label{eq:params}
\end{equation}
where $t_c$ and $\phi_c$ are the time and phase at coalescence, $\mathcal{M}_c$ is the chirp mass, $\eta$ the symmetric mass ratio, $\chi_{1,2}$ the aligned dimensionless spins, $\kappa_s$ the symmetric spin-induced quadrupole coefficient (with $\kappa_a=0$), and $e_0$ the initial eccentricity. The luminosity distance and the angular parameters are held fixed. All errors quoted below are marginalized over the full vector~\eqref{eq:params}.
The integral in Eq.~\eqref{eqn6} is terminated at twice the orbital frequency of the innermost stable circular orbit (ISCO) for a \textit{spinning} compact binary, using the fits given in \cite{Buonanno:2009zt}. For the lower frequency cutoff, the sensitivity of the detector is determined by the function \(S_{n}(f)\). We focus on parameter estimation for two different third-generation gravitational-wave detectors: Cosmic Explorer (CE) \cite{Reitze:2019iox} and Einstein Telescope (ET-D) \cite{Punturo:2010zz,Maggiore:2019uih}. We compare the performance of CE and ET-D (hereafter ET) throughout. For CE (ET) the lower frequency cutoff is chosen to be 5 Hz (1 Hz), which corresponds to \(f_{\text{low}}\) in the integral given in Eq.~\eqref{eqn6}. In the single-event and spin-plane analyses of Sec.~\ref{sec:single_event_results} the initial eccentricity $e_0$ is defined at this lower cutoff frequency, i.e.\ at $5\,\mathrm{Hz}$ for CE and $1\,\mathrm{Hz}$ for ET; the population study uses a common reference frequency (Sec.~\ref{sec:population_synthesis}). Advanced LIGO is used only in the template-bank study of Sec.~\ref{sec:temp_bank}.

\section{Results}\label{sec:results}

\subsection{Single-Event Forecasts: Impact of Eccentricity on SIQM Measurements}\label{sec:single_event_results}

We assume the eccentric binary to be at a luminosity distance of $400\,\mathrm{Mpc}$, with total mass ranging from $M_{\rm tot}=10$ to $100\,M_\odot$, and compare Cosmic Explorer (CE, $f_{\rm low}=5\,\mathrm{Hz}$) with Einstein Telescope (ET, $f_{\rm low}=1\,\mathrm{Hz}$) across this full range. We first examine the mass dependence of $\Delta\kappa_s$ and $\Delta e_0$ at a fixed, representative spin configuration. We then map the errors across the full $\chi_1$--$\chi_2$ spin plane, allowing us to explore how spin magnitude and alignment affect the measurability of $\kappa_s$ and the eccentricity parameter, and to quantify their correlation.

\begin{figure*}[htbp]
  \centering
  \includegraphics[width=\textwidth]{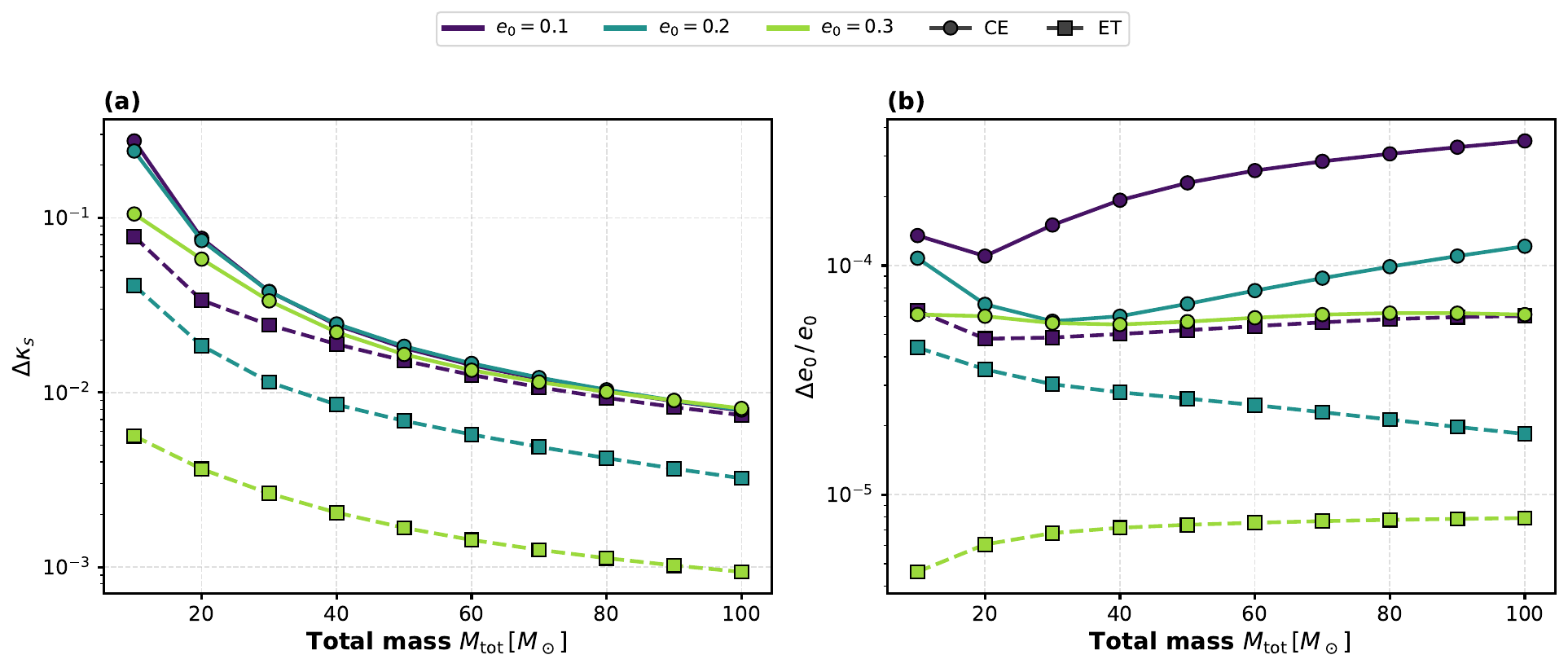}
  \caption{\justifying\textbf{Single-event uncertainties as a function of total mass.}
           (a)~$\Delta\kappa_s$ and (b)~fractional eccentricity uncertainty $\Delta e_0/e_0$ for
           $e_0=0.1$, $0.2$, and $0.3$ (colors). Circles with solid lines denote CE; squares with
           dashed lines denote ET. The binary is optimally oriented at $400\,\mathrm{Mpc}$ with mass
           ratio $q=1.25$ and aligned spins $\chi_1=0.9$, $\chi_2=0.8$ (heavier and lighter component,
           respectively); both vertical axes are logarithmic. ET gives the smaller $\Delta\kappa_s$ at
           every mass and eccentricity shown, with $\Delta\kappa_s^{\rm CE}/\Delta\kappa_s^{\rm ET}$
           spanning $1.1$--$3.5$ at $e_0=0.1$, $2.5$--$5.9$ at $e_0=0.2$, and $8.6$--$19$ at $e_0=0.3$.
           The same ordering holds for $\Delta e_0/e_0$, with CE/ET ratios of $2.1$--$5.8$, $1.9$--$6.6$,
           and $7.7$--$13$, respectively.}
  \label{fig:single_event}
\end{figure*}

Figure~\ref{fig:single_event}(a) shows the $1\sigma$ uncertainty on $\kappa_s$ as a function of total mass for $e_0=0.1$, $0.2$, and $0.3$ at fixed spins $\chi_1=0.9$, $\chi_2=0.8$. For both detectors and all three eccentricities, $\Delta\kappa_s$ decreases monotonically with mass, following the growth of the SNR. ET yields the smaller $\Delta\kappa_s$ at every mass for all three eccentricities. The ratio $\Delta\kappa_s^{\rm CE}/\Delta\kappa_s^{\rm ET}$ falls with mass, from $3.5$ at $10\,M_\odot$ to $1.1$ at $100\,M_\odot$ for $e_0=0.1$ (near parity at high mass), from $5.9$ to $2.5$ for $e_0=0.2$, and from $18.7$ to $8.6$ for $e_0=0.3$. The ET advantage therefore grows with $e_0$, and for $e_0\geq0.2$ the ratio stays above $2.4$ across the full mass range. This behavior is consistent with the eccentric correction benefiting most the detector whose low-frequency sensitivity captures more of the eccentric inspiral. These results use the $\mathcal{O}(e_0^8)$-corrected waveform of Sridhar~et~al.~\cite{Omkar2025}.\footnote{We use the expansion to eighth order in the initial eccentricity $e_0$, as derived in Ref.~\cite{Omkar2025}. This order is necessary to accurately capture the eccentricity corrections in the moderate-eccentricity regime considered here ($e_0=0.1$--$0.3$).}

Figure~\ref{fig:single_event}(b) shows the fractional uncertainty $\Delta e_0/e_0$ for the same systems. At fixed $e_0$ the CE/ET ratio is identical for $\Delta e_0$ and $\Delta e_0/e_0$. ET again gives the smaller error at every mass and eccentricity, by factors of $2.1$--$5.8$ ($e_0=0.1$), $1.9$--$6.6$ ($e_0=0.2$), and $7.7$--$13$ ($e_0=0.3$). At $10\,M_\odot$, ET reaches $\Delta e_0/e_0=6.4\times10^{-5}$, $4.4\times10^{-5}$, and $4.6\times10^{-6}$ for $e_0=0.1$, $0.2$, and $0.3$, compared with $1.4\times10^{-4}$, $1.1\times10^{-4}$, and $6.1\times10^{-5}$ for CE. The mass dependence is not uniform across eccentricities. For CE at $e_0=0.1$ and $0.2$, the error decreases from $10\,M_\odot$ to a minimum near $20\,M_\odot$ and $30\,M_\odot$, respectively, and then rises by factors of $\sim3.2$ and $\sim2.1$ out to $100\,M_\odot$; at $e_0=0.3$ it varies by $\lesssim12\%$ across the mass range. For ET, the error decreases monotonically with mass only at $e_0=0.2$ (from $4.4\times10^{-5}$ to $1.8\times10^{-5}$). At $e_0=0.1$ it dips between $10$ and $20\,M_\odot$ and then rises by $\sim26\%$ out to $100\,M_\odot$, and at $e_0=0.3$ it increases with mass, from $4.6\times10^{-6}$ to $7.9\times10^{-6}$.

Because Fig.~\ref{fig:single_event} fixes the spins at $(\chi_1,\chi_2)=(0.9,0.8)$ -- a high-spin configuration chosen to maximize the SIQM contribution -- they cannot address how strongly $\Delta\kappa_s$ and $\Delta e_0$ depend on spin alignment itself, or whether the ET/CE hierarchy established here persists across the full spin plane. We address this next by fixing the total mass and scanning $(\chi_1,\chi_2)$ directly.

\paragraph{Spin-plane dependence at $M=10\,M_\odot$}\footnote{The spin-plane study uses the $\mathcal{O}(e_0^2)$ truncation of the eccentric phasing rather than the $\mathcal{O}(e_0^8)$ expansion of Fig.~\ref{fig:single_event}. The run requires several hundred Fisher evaluations per detector and eccentricity, and the higher-order expressions make it computationally expensive. Because the aim is a qualitative picture of how the errors depend on spin alignment, we adopt the lower order. Each panel uses 400 random $(\chi_1,\chi_2)$ pairs.}
\begin{figure*}[t]
    \centering
    \includegraphics[width=\textwidth]{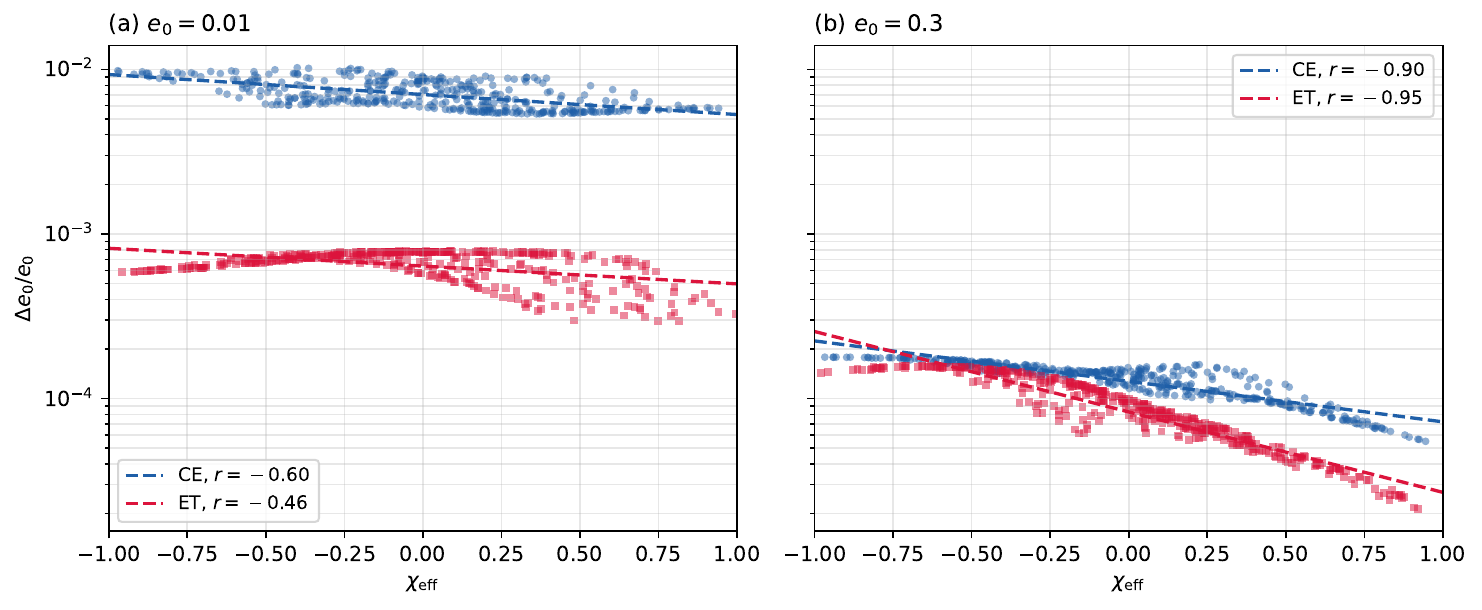}
    \caption{\justifying\textbf{Fractional eccentricity uncertainty versus effective spin at $M=10\,M_\odot$.}
             $\Delta e_0/e_0$ as a function of $\chi_{\rm eff}$ for 400 random spin pairs at (a)~$e_0=0.01$ and (b)~$e_0=0.3$ (CE blue circles, ET crimson squares); the vertical axis is shared and logarithmic. Dashed lines are log-linear fits, with Pearson coefficients $r$ between $\log_{10}(\Delta e_0/e_0)$ and $\chi_{\rm eff}$ in the legends. Both detectors improve with eccentricity, CE more strongly, and the correlation with $\chi_{\rm eff}$ strengthens for both. Computed at $\mathcal{O}(e_0^2)$.}
    \label{fig:m10}
\end{figure*}

We sample the spin plane at fixed $M=10\,M_\odot$ and examine how the errors depend on the effective spin $\chi_{\rm eff}=(q\chi_1+\chi_2)/(1+q)$. Figure~\ref{fig:m10} shows $\Delta e_0/e_0$ versus $\chi_{\rm eff}$ at $e_0=0.01$ and $0.3$. At fixed $e_0$ the detector ratio is the same for $\Delta e_0$ and $\Delta e_0/e_0$, but the relative error is the meaningful quantity for comparing the two eccentricities.

At low eccentricity, ET reaches a median $\Delta e_0/e_0=7.0\times10^{-4}$, a factor of $\simeq9.7$ below CE ($6.8\times10^{-3}$), reflecting ET's access to the long low-frequency inspiral in which eccentricity information accumulates. The Pearson coefficient $r$ between $\log_{10}(\Delta e_0/e_0)$ and $\chi_{\rm eff}$ is moderate for both detectors, $r=-0.46$ (ET) and $r=-0.60$ (CE).

At $e_0=0.3$ both detectors improve. The CE median falls by a factor of $\simeq48$ (to $1.4\times10^{-4}$) and the ET median by $\simeq8.4$ (to $8.3\times10^{-5}$), so the CE/ET ratio narrows from $9.7$ to $1.7$. The anticorrelation with $\chi_{\rm eff}$ strengthens sharply for both (CE: $r=-0.60\to-0.90$; ET: $r=-0.46\to-0.95$), and the scatter about the fit shrinks, so that at high eccentricity $\chi_{\rm eff}$ accounts for most of the variation in the error ($r^2\simeq0.8$--$0.9$).

The ratio $\mathcal{R}_{e_0} = \Delta e_0(\chi_{\rm eff}<0)/\Delta e_0(\chi_{\rm eff}>0)$ quantifies the orbital hang-up effect\footnote{The orbital hang-up effect refers to the increase in the number of inspiral cycles for aligned spins (positive $\chi_{\rm eff}$) compared to anti-aligned spins, which improves the measurement precision of parameters such as $e_0$. The ratio $\mathcal{R}_{e_0}>1$ indicates that aligned spins yield smaller $\Delta e_0$ than anti-aligned spins.}. For CE, $\mathcal{R}_{e_0}=1.35$ at $e_0=0.01$ and $1.43$ at $e_0=0.3$. For ET, $\mathcal{R}_{e_0}$ rises from $1.12$ to $2.00$, so the hang-up effect strengthens markedly at high eccentricity.

The error on $\kappa_s$ is nearly independent of spin alignment: $|r|\lesssim0.14$ for both detectors at both eccentricities. This is consistent with $\kappa_s$ coupling to the spin magnitudes ($\chi^2$) rather than to the signed effective spin, and with the weak alignment dependence at fixed mass in Fig.~\ref{fig:single_event}(a). Eccentricity nevertheless lowers the median $\Delta\kappa_s$ for both detectors, by a factor of $4.3$ for ET ($3.34\times10^{-1}\to7.75\times10^{-2}$) and $1.7$ for CE ($5.13\times10^{-1}\to3.09\times10^{-1}$), in line with the growing ET/CE gap in Fig.~\ref{fig:single_event}(a).

\paragraph{Comparison at $M=90\,M_\odot$}

The corresponding analysis at $M=90\,M_\odot$ (Fig.~\ref{fig:m90_appendix} and Table~\ref{tab:m90_appendix}, Appendix~A) gives the same qualitative picture. Both detectors improve with eccentricity in $\Delta e_0/e_0$ (CE by $\simeq350$, ET by $\simeq28$), and the CE/ET ratio narrows from $\simeq24$ at $e_0=0.01$ to $\simeq1.9$ at $e_0=0.3$. The $\chi_{\rm eff}$ dependence of $\Delta e_0$ again strengthens with eccentricity (ET: $r=0.26\to-0.99$, with $\mathcal{R}_{e_0}$ rising from $0.93$ to $1.82$), while $\Delta\kappa_s$ remains insensitive to spin alignment ($|r|<0.13$). The single-event forecasts therefore indicate that the ET/CE hierarchy for $\Delta\kappa_s$ is set primarily by eccentricity and mass, and that spin alignment matters mainly through its mass- and eccentricity-dependent influence on $\Delta e_0$.

\subsection{Population-Level Comparisons: Cosmic Explorer vs. Einstein Telescope}
\label{sec:population_synthesis}

\paragraph{Methodology}

To simulate the population of detectable gravitational-wave sources across a range of cosmological redshifts, we draw sources according to the astrophysically motivated probability density
\begin{equation}
p(z) \propto \mathcal{R}(z)\,\frac{dV_c}{dz},
\label{eqn:pop_pz}
\end{equation}
where $\mathcal{R}(z)$ is the star-formation-rate-weighted merger-rate evolution and $dV_c/dz$ is the comoving volume element, evaluated using the $D_L(z)$ relation of Sec.~\ref{sec:PE} with $H_0=67.66\,\mathrm{km\,s^{-1}\,Mpc^{-1}}$ and $\Omega_m=0.315$. We precompute the cumulative distribution function of Eq.~\eqref{eqn:pop_pz} to $z_{\rm max}=1$ and draw redshifts via inverse-transform sampling. Component masses are drawn either uniformly on $[5,20]\,M_\odot$ per component (Uniform configuration) or with $m_1\sim p(m_1)\propto m_1^{-2.3}$ and $m_2$ uniform on the same range (Power-law configuration); spins are drawn independently and uniformly on $\chi_{1,2}\in[0.8,0.95]$, and initial eccentricity uniformly on $e_0\in[0.1,0.5]$, defined at a reference frequency of $10\,\mathrm{Hz}$ for both detectors. The octupole coefficient is fixed to its Kerr value, $\lambda_s=1$ (with $\lambda_a=\kappa_a=0$), and $\kappa_s=1$ at the injection point.

For each drawn source we compute the full Fisher matrix of Sec.~\ref{sec:PE} directly at the source's true luminosity distance $D_L(z)$, and retain the source if $\rho\geq\rho_{\rm th}=10$. The uncertainty $\Delta{\kappa_s}$ (and $\Delta{e_0}$) is extracted as the square root of the corresponding diagonal element of the inverse Fisher matrix.

The eccentric phasing is an expansion in $e_0$ defined at $f_0=10\,\mathrm{Hz}$, with effective expansion parameter $e(f)\simeq e_0\,(f_0/f)^{19/18}$ at frequency $f$. At the lower cutoff of each detector this gives $e(f_{\rm low})\simeq11.4\,e_0$ for ET ($1\,\mathrm{Hz}$) and $\simeq2.1\,e_0$ for CE ($5\,\mathrm{Hz}$). The expansion parameter therefore exceeds unity at the low-frequency edge of the band for $e_0\gtrsim0.09$ in ET and $e_0\gtrsim0.48$ in CE. The population results, and the ET results in particular, thus extrapolate the truncated expansion beyond its regime of validity and indicate trends rather than quantitative forecasts.

\paragraph{Exploratory population results for 1000 binaries}

\begin{figure*}[t]
    \centering
    \includegraphics[width=\textwidth]{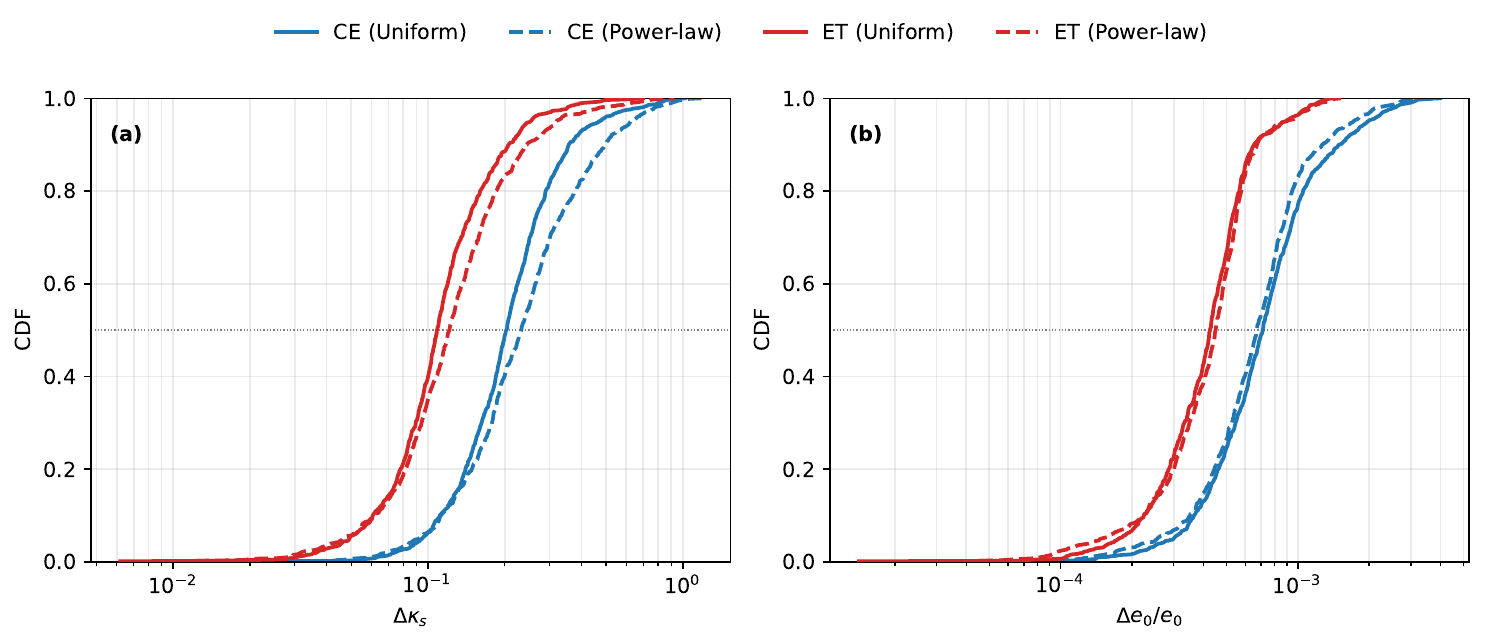}
    \caption{\justifying\textbf{Population CDFs for $N=1000$ binaries, $\mathcal{O}(e_0^4)$ waveform.} (a) $\Delta \kappa_s$ and (b) fractional eccentricity uncertainty $\Delta e_0/e_0$, evaluated per source with its injected $e_0$, for CE (Uniform: solid blue; Power-law: dashed blue) and ET (Uniform: solid red; Power-law: dashed red) component-mass distributions.}
    \label{fig:pop_cdf_n1000}
\end{figure*}

We draw $N=1000$ binaries for each detector and mass distribution, using the $\mathcal{O}(e_0^4)$ waveform. Figure~\ref{fig:pop_cdf_n1000}(a) shows the CDF of $\Delta{\kappa_s}$. The median is $0.109$ (Uniform) and $0.121$ (Power-law) for ET, and $0.203$ and $0.232$ for CE, so ET is tighter by a factor of $\sim1.9$ for either prior. Eccentric signatures are most pronounced at low frequency, and ET's extended low-frequency reach lets it exploit the early-inspiral eccentric dynamics that CE's higher-frequency sensitivity partially misses, consistent with the single-event ordering of Sec.~\ref{sec:single_event_results}.

Panel~(b) shows the fractional uncertainty $\Delta e_0/e_0$. The median is $4.3\times10^{-4}$ (Uniform) and $4.5\times10^{-4}$ (Power-law) for ET, and $7.1\times10^{-4}$ and $6.7\times10^{-4}$ for CE, so ET improves the eccentricity measurement by a factor of $1.5$--$1.7$. The central $68\%$ of the Uniform distributions spans $2.6$--$5.9\times10^{-4}$ for ET and $4.3$--$12\times10^{-4}$ for CE; the CE distribution has the longer high-error tail. The ET/CE hierarchy is unchanged by the mass prior in both $\Delta{\kappa_s}$ and $\Delta{e_0}/e_0$. The prior affects $\Delta\kappa_s$ at the $12$--$14\%$ level, with a heavier high-error tail for the Power-law CE case, and $\Delta e_0/e_0$ at the $\lesssim6\%$ level, where the Uniform and Power-law curves nearly coincide.

\paragraph{Astrophysical Implications}
The sensitivity-weighted sampling to $z_{\rm max}=1$ captures the bulk of the 3G-detectable population at the masses considered here. If the ET median $\Delta\kappa_s\sim0.1$ were realized, it would be precise enough to exclude the Kerr value $\kappa_s=1$ from the neutron-star range ($\kappa_s\sim2\text{--}14$, Sec.~\ref{sec:waveformmodel}) at high significance, and to constrain the broader range spanned by boson-star and gravastar models \cite{Ryan:1996nk,Adam:2022pdb,Uchikata:2016qku}. Given the validity limitation of the truncated eccentric expansion noted above, we regard these numbers as indicative. Translated into bounds on rotating exotic-compact-object parameters \cite{Krishnendu:2025rud}, they suggest that eccentric binaries may yield SIQM constraints that complement circular-orbit tests.

\section{Importance of including eccentricity in SIQM templates}
\label{sec:temp_bank}
\begin{figure*}[t]
    \centering
    \includegraphics[width=\linewidth]{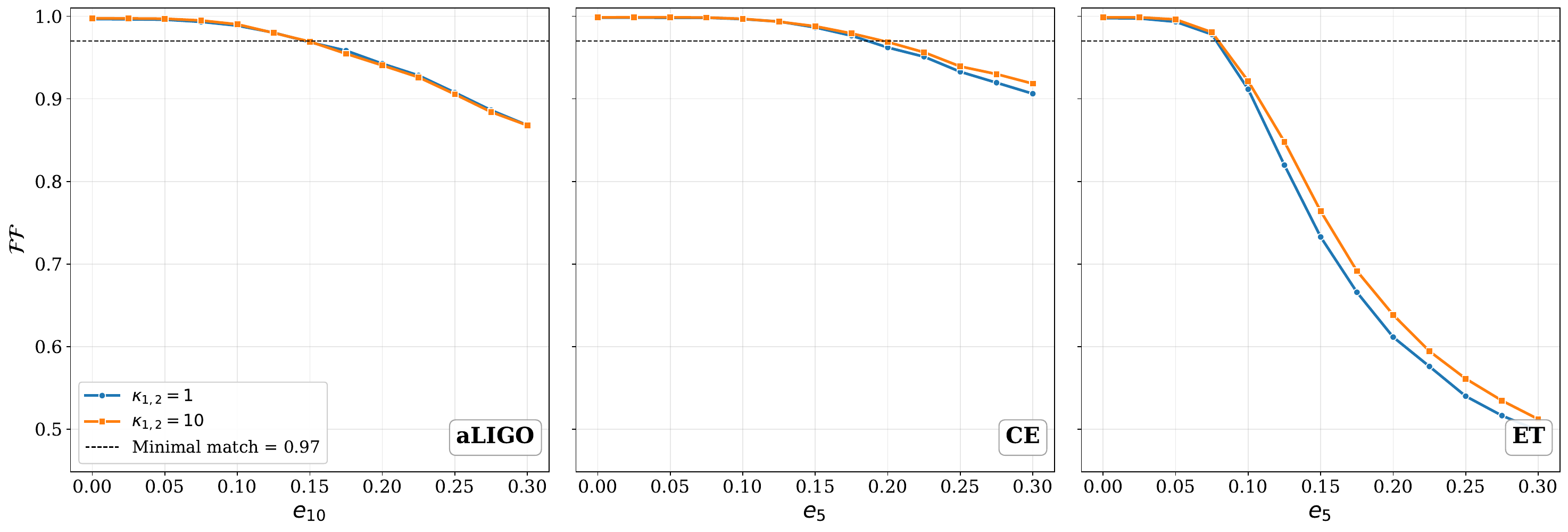}\hfill
    \includegraphics[width=\linewidth]{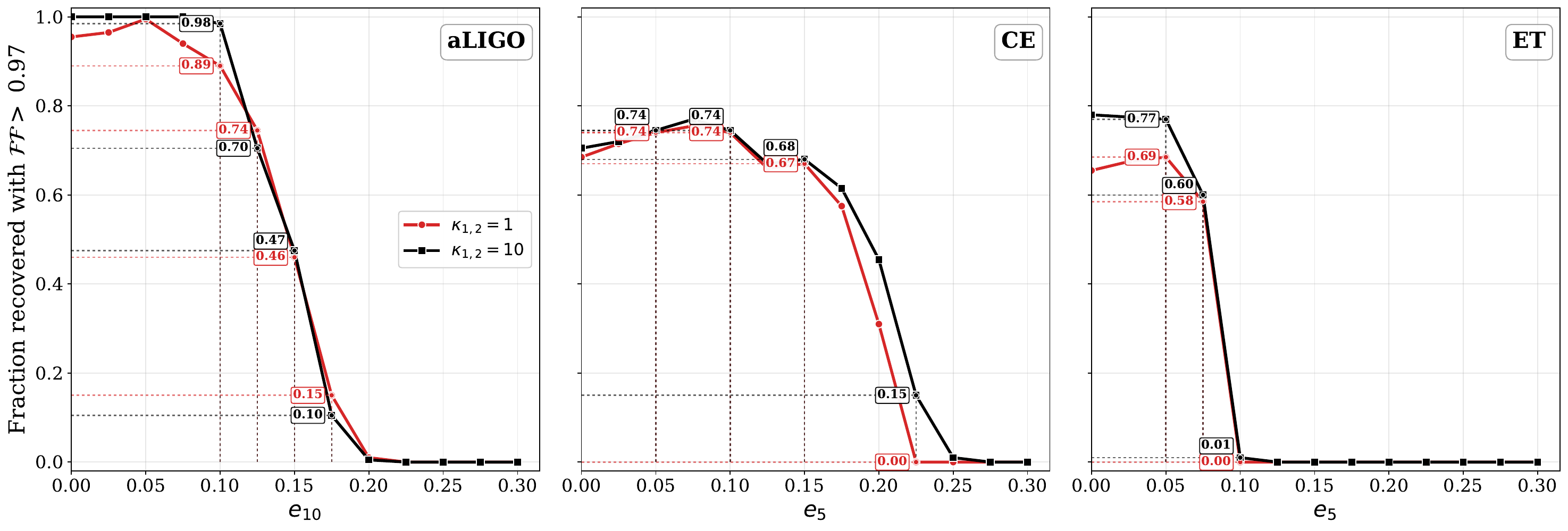}\hfill
    \caption{\justifying The effect of eccentricity on $\mathcal{FF}$ computation using circular templates for fixed $\kappa_{1,2}$ is shown here. Top row: $\mathcal{FF}$ is plotted against initial eccentricity for aLIGO (left), CE (middle), and ET (right). The horizontal dashed black line denotes the minimal match of $0.97$ set to perform the computation. The $\mathcal{FF}$ decreases with increasing eccentricity for all three detectors, ET showing the largest drop. This can be attributed to ET being much more sensitive at lower frequencies resulting in capturing more eccentric GW cycles compared to CE and aLIGO.
    Bottom row: The fraction of eccentric aligned-spin injections recovered with $\mathcal{FF}>0.97$ by circular aligned-spin templates as a function of initial eccentricity is shown for aLIGO (left), CE (middle), ET (right), respectively. Same as the figure from top panel, the drop in fraction of injections recovered with $\mathcal{FF}> 0.97$ is sharpest for ET compared to CE and aLIGO.}
    \label{fig:ff_and_frac_recovered}
\end{figure*}
To assess the importance of inclusion of eccentricity in the template banks used in traditional LVK search pipelines, in this section, we perform fitting factor $(\mathcal{FF})$~\citep{Apostolatos:1995pj} computation using the inspiral-only, eccentric and aligned-spin waveform model in frequency domain presented in Sec.~III C of \citep{Omkar2025}. $\mathcal{FF}$ is defined as the
maximum fraction of the optimal SNR of a given signal
that a given template bank is capable of recovering (see Sec. 2 of \citep{Kacanja:2024pjh} for details on $\mathcal{FF}$). The $\mathcal{FF}$ study is carried out using advanced LIGO (aLIGO) and proposed third-generation (3G) detectors such as CE and ET PSDs.\\
\subsection{Template banks}
\label{subsec:tb}
{\red We use the brute force stochastic method algorithm \texttt{pycbc\_brute\_bank} of PyCBC~\citep{alex_nitz_2020_4134752} to create a four-dimensional, quasi-circular, aligned-spin template bank fixing the SIQM parameters $\kappa_{1,2}$ to 1 (the Kerr BH value) and 10, respectively. The individual component masses $(m_{1,2})$ and $z-$component of the dimensional spins $(\chi_{1z},\chi_{2z})$ of the compact objects are varied in a range of values for aLIGO, CE and ET to construct the bank. The details of the parameters and range of values are outlined in Table~\ref{tab:tb_params}. We use a lower-cutoff frequency $(f_{\rm low})$ of 10 Hz and use \texttt{aLIGOZeroDetLowPower} design sensitivity curve for aLIGO. Besides, \texttt{CosmicExplorerP1600143} and \texttt{EinsteinTelescopeP1600143} sensitivity curves are used with $f_{\rm low}$ set to 5 Hz for CE and ET, respectively. Further, we choose a design minimal match of $0.97$ (or equivalently 97\%) to construct the bank. Additionally, since an inspiral-only waveform is used to generate the bank, we restrict the upper limit of total mass to $60 M_{\odot}$ and $80 M_{\odot}$ for aLIGO and CE, ET detectors, respectively.\footnote{\red The restriction to $M \leq 80\,M_\odot$ and $q \leq 3$, and the
$20\,M_\odot$ lower value of the component masses adopted for CE and ET detectors, is imposed for computational tractability rather than on astrophysical grounds.  At $f_{\rm low} = 5~$Hz the in-band duration
grows steeply toward low masses: a $\left(20+20\right)\,M_\odot$ binary is
$77~$seconds long, whereas a $\left(10+10\right)\,M_\odot$ binary is approximately of
$245~$seconds. The number of templates required to cover the whole parameter space with a minimal match of 0.97 and the cost of each \red individual match calculation
grows with increasing ranges of parameters. This restriction is \red conservative for the purposes of this work.  The excluded low total mass and high mass-ratio corner of the parameter space
accumulates more waveform cycles in band, and would therefore
expected to exhibit a larger degradation in $\mathcal{FF}$ in the absence of eccentricity in the templates than the region studied here.
}}

\begin{table}[t]
\caption{\justifying Parameter ranges spanned by the quasi-circular, aligned-spin
template banks constructed for each detector configuration.  Component masses $(m_{1,2})$ are
quoted in the detector frame; $\chi_{1z}$ and $\chi_{2z}$ are the $z$-components of the dimensionless spin vectors of compact objects. Two different set of banks are generated for each detector by fixing the SIQM parameters $\kappa_{1,2}$ to 1 (the Kerr BH limit) and 10.}
\label{tab:tb_params}
\begin{ruledtabular}
\begin{tabular}{lccc}
Parameter & aLIGO & CE & ET \\
\colrule
$m_1,\,m_2$ $(M_\odot)$        & $[10,\,40]$      & $[20,\,60]$      & $[20,\,60]$      \\
$M = m_1 + m_2$ $(M_\odot)$    & $[20, 60]$        & $[40, 80]$        & $[40, 80]$        \\
$q = m_1/m_2$                  & $[1, 4]$         & $[1, 3]$         & $[1, 3]$         \\
$\chi_{1z},\,\chi_{2z}$        & $[-0.8,\,0.8]$   & $[-0.8,\,0.8]$   & $[-0.8,\,0.8]$   \\
\end{tabular}
\end{ruledtabular}
\end{table}

\subsection{Injection set and fitting factor}
\label{subsec:inj_ff}
{\red 
The injected signals are generated using the same waveform model that has been used to create the template banks (and presented in Sec. III C of \citep{Omkar2025}), with the value of $\kappa_{1,2}$ fixed in both the injection and templates used for the $\mathcal{FF}$ computation. This is done to ensure that the
recovered mismatch isolates the loss due to the absence of eccentricity in the template bank and is not contaminated by waveform systematics or
SIQM parameter value mismatch effects. Component masses and $z-$component of dimensional spins are drawn uniformly within the bank's total mass and dimensionless spin bounds as given in Table~\ref{tab:tb_params}. For each value of the initial eccentricity we draw $200$ injections, on a grid of eccentricity $ \in [0, 0.3]$ in steps of $0.025$, giving $2600$ injections per detector for a fixed value of $\kappa_{1,2}$. We define the eccentricity at $10\,\mathrm{Hz}$ for aLIGO ($e_{10}$) and at $5\,\mathrm{Hz}$ for both CE and ET ($e_{5}$), the respective lower cutoff frequencies of the banks; this distinction is important when comparing detectors and we return to it in Sec.~\ref{subsec:ff_implication}. Throughout the FF study, we regard an injection is recovered when the $\mathcal{FF} > \text{minimal match}=0.97$ for that particular injection with the circular template.}
\subsection{Implications}
\label{subsec:ff_implication}
{\red The top panel of Figure~\ref{fig:ff_and_frac_recovered} shows the median  $\mathcal{FF}$ as a function of
initial eccentricity for aLIGO (left), CE (middle), and ET (right) detectors for two distinct values of $\kappa_{1,2}=1,\, 10$ denoted by blue and orange, respectively. We can observe that the $\mathcal{FF}$ decreases with increasing initial eccentricity for all detectors. This is expected because eccentricity induced oscillations in the injected signals will become more prominent with increasing eccentricity and since the template is circular and devoid of these oscillations, this will result in higher accumulated dephasing (and hence higher mismatch) between the two with increasing eccentricity. Additionally, the drop in $\mathcal{FF}$ is steeper for ET compared to CE and aLIGO. This is due to the fact that ET is much more sensitive \footnote{As can be seen from Figure 1 of Ref.~\citep{Divyajyoti:2021uty}, ET is more than one and three orders of magnitude more sensitive compared to CE and aLIGO in the frequency range $5-10$ Hz.} in the low frequency range $5-10$ Hz compared to CE and aLIGO. This causes ET to capture more eccentric cycles compared to CE and aLIGO, resulting in larger drop in SNR (and hence $\mathcal{FF}$) while recovering it using circular templates. The bottom panel of Figure~\ref{fig:ff_and_frac_recovered} shows the corresponding fraction of eccentric injections recovered above threshold $0.97$ by the circular template bank for two different fixed values of $\kappa_{1,2}$ denoted by different colors. The two figures in both the panels are complementary to each other. The former measures the typical loss of SNR as a function of initial eccentricity, the latter addresses the detection related question of what fraction of a eccentric population is still found using circular templates for fixed SIQM parameter values. Note that a 10\% loss in
SNR $(\text{equivalently}, \mathcal{FF}=0.9)$ results in approximately a 27\% reduction in overall detection volume (or observation rate, provided sources are uniformly distributed in comoving volume). Below we go further into the details of these figures for each detector.
\paragraph{Advanced LIGO:} The quasi-circular bank is able to recover the threshold value of 0.97 for $e_{10} < 0.15$ (top panel of Figure~\ref{fig:ff_and_frac_recovered}). The median $\mathcal{FF}$ crosses the threshold value at $e_{10}=0.15$. Beyond that, the drop in $\mathcal{FF}$ is rapid. For instance,  $\simeq90\%$ of the injections have been recovered with $\mathcal{FF}>0.97$ for both $\kappa_{1,2}$ cases for $e_{10}\lesssim0.10$ (bottom panel of Figure~\ref{fig:ff_and_frac_recovered}). However, this fraction decreases significantly and falls from 0.89 for $\kappa_{1,2}=1$ (correspondingly, 0.98 for $\kappa_{1,2}=10$) at $e_{10}=0.10$ to 0.46 (0.47) at $e_{10}=0.15$ and to 0.15 (0.10) at $e_{10}=0.175$. At $e_{10}=0.2$ no fractions of injections are recovered with $\mathcal{FF}>0.97$ for both $\kappa_{1,2}$ cases. 
\paragraph{Cosmic Explorer:}
The behavior of $\mathcal{FF}$ for CE is qualitatively similar but slightly shifted compared to aLIGO. The median $\mathcal{FF}$ crosses threshold at $e_{5}\sim 0.19$ (top panel of Figure~\ref{fig:ff_and_frac_recovered}) and recovered fraction falling from 0.74 at $e_{5}=0.10$ to $\sim 0.68$ at $e_5 = 0.15$ for both $\kappa_{1,2}$ cases (bottom panel of Figure~\ref{fig:ff_and_frac_recovered}). At $e_{5}=0.20$ no injections are recovered for $\kappa_{1,2}=1$ and for $\kappa_{1,2}=10$ it drops to 0.15. For largest considered $e_{5}=0.3$, the $\mathcal{FF}\gtrsim 0.90$ for both $\kappa_{1,2}$ cases.
\paragraph{Einstein Telescope:}
ET is less tolerant of eccentricity than either of
the other configurations. The median $\mathcal{FF}$ crosses threshold at $e_{5} \simeq 0.076$ (top panel of Figure~\ref{fig:ff_and_frac_recovered}), and the recovered fraction drastically reduces to zero over a single grid interval, from $\simeq 0.6$ at $e_5 = 0.075$ to $\leq 0.01$ at $e_{5} = 0.10$ for both $\kappa_{1,2}$ cases (bottom panel of Figure~\ref{fig:ff_and_frac_recovered}). The loss in SNR is also far more severe than for the other detectors: the median $\mathcal{FF}$ falls to  $\simeq 0.50$ at $e_5 = 0.3$, corresponding to the recovery of only about half of the available SNR by the best quasi-circular template.

The $\mathcal{FF}$ values can't be compared between aLIGO and CE, ET because the eccentricities for the former is defined at 10 Hz and for the latter two a bit earlier at 5 Hz. Since eccentricity decays over time due to GW emission, we can use the leading order solutions of eccentricity evolution equation: $e\left(f\right) = e_0 \left(\frac{f_0}{f}\right)^{19/18}$~\citep{Peters:1964zz,Moore:2016gqo} to map the $e_5$ values to $e_{10}$\footnote{Here, $e_0$ is the eccentricity defined at initial frequency $f_0$.}. For example, $e_5 = 0.1$ for CE and ET will correspond to $e_{10}\simeq0.05$ for the same binary. Combining this fact with the bottom panel of Figure~\ref{fig:ff_and_frac_recovered},  $\simeq90\%$ or more of the binaries with $e_{10}\leq 0.05$ for aLIGO are recovered with $\mathcal{FF}>0.97$ for both $\kappa_{1,2}$ cases using a quasi-circular template. However, the same population of binaries when observed via CE and ET detectors, the fraction drastically reduces to 0.74 and 0.0, respectively. All these observations point to the fact that the eccentric binaries to
which a quasi-circular template-based search remains effectual is not a
fixed number but a strong function of the detector’s low-frequency sensitivity. For aLIGO the quasi-circular bank is adequate for the eccentricities expected from the bulk of field binaries, and only dynamically formed systems for which $e_{10} \gtrsim 0.1$ at detection would be significantly penalized. For 3G
detectors and for ET in particular, the tolerance is far narrower. A quasi-circular bank
loses more than half of a population with $e_{5} \sim 0.08$, and
recovers only about half of the available SNR for $e_5 = 0.3$. Highly accurate eccentric template banks will therefore be considerably more important in the 3G era than they are at present.
}
\section{Conclusion}\label{sec:conclusion}

In this work we extend the symmetric spin-induced quadrupole-moment (SIQM) test, $\kappa_s$, to eccentric, aligned-spin compact binaries using the spinning-eccentric post-Newtonian phasing model of~\citep{Omkar2025}, and forecast its measurability with third-generation (3G) detectors. Unlike earlier treatments in which eccentricity enters only additively~\cite{Krishnendu:2026eccSIQM}, this model couples $\kappa_s$ to both the circular and eccentric parts of the phase, allowing us to assess whether eccentricity degrades or strengthens the SIQM null test rather than assuming the former.

Our single-event Fisher-matrix forecasts (Sec.~\ref{sec:single_event_results}) show that the size of the eccentric contribution depends on total mass, detector, and eccentricity. At fixed high spin ($\chi_1=0.9$, $\chi_2=0.8$), ET yields a smaller $\Delta\kappa_s$ than CE at every mass considered, with $\Delta\kappa_s^{\rm CE}/\Delta\kappa_s^{\rm ET}$ rising from $1.1$--$3.5$ at $e_0=0.1$ to $8.6$--$19$ at $e_0=0.3$, and ET yields a smaller fractional eccentricity uncertainty $\Delta e_0/e_0$ than CE by a factor of $2$--$13$ over the same range. The spin-plane analysis at $M=10\,M_\odot$ and $M=90\,M_\odot$ shows that this ET advantage is driven almost entirely by mass and eccentricity: $\Delta\kappa_s$ depends only weakly on spin alignment at either mass, consistent with its quadratic coupling to spin magnitude, while $\Delta e_0$ develops a strong, eccentricity-dependent hang-up correlation with $\chi_{\rm eff}$ at both masses. Together these results indicate that eccentricity, rather than acting purely as a nuisance parameter to be marginalized over, supplies complementary phase information that can improve the discriminating power of the SIQM test -- provided the detector retains sufficient low-frequency sensitivity to capture it.

A similar ordering is found at the population level (Sec.~\ref{sec:population_synthesis}). In an exploratory population study with a truncated eccentric expansion, for a sensitivity-weighted population out to $z=1$ ($N=1000$ binaries per configuration), the median $\Delta\kappa_s$ is $0.11$--$0.12$ for ET and $0.20$--$0.23$ for CE, and the median $\Delta e_0/e_0$ is $\sim4$--$5\times10^{-4}$ for ET and $\sim7\times10^{-4}$ for CE. The ET advantage is insensitive to the assumed component-mass distribution (Uniform vs.\ Power-law), but, because the expansion parameter exceeds unity at the low-frequency edge of the ET band for most of the sampled $e_0$, we regard the absolute values as indicative.

Finally, in Sec.~\ref{sec:temp_bank} we quantify the importance of inclusion of eccentricity in templates where SIQM parameters are set to Kerr BH $(\kappa_{1,2}=1)$ and non-Kerr BH $(\kappa_{1,2}=10)$ values. We created quasi-circular aligned-spin templates and computed $\mathcal{FF}$ using the eccentric aligned-spin injections for two different SIQM parameters fixed in both injection and template for aLIGO, CE, and ET detectors. The top row of Figure~\ref{fig:ff_and_frac_recovered} shows the change in $\mathcal{FF}$ as a function of initial eccentricity and the bottom row quantifies the corresponding fraction of injections being recovered by the quasi-circular template bank with $\mathcal{FF}>0.97$. Both the figures point to the fact that $\mathcal{FF}$ significantly decreases with increasing eccentricity, ET showing the largest drop. This indicates the sheer importance of including effects due to eccentricity in the templates used for searches, especially for 3G detectors.

Several caveats apply to the results presented here. First, our Fisher-matrix forecasts rely on the standard high-SNR, Gaussian-noise approximation, which is expected to be reliable for the SNRs considered but does not capture non-Gaussianities or prior information that a full Bayesian analysis would include. Second, we have restricted the analysis to aligned spins; spin precession, which can itself mimic or mask eccentricity- and quadrupole-induced phase modulations, is not addressed here. Third, in the single-event and spin-plane analyses $e_0$ is defined at each detector's lower cutoff frequency, so the CE and ET comparisons at equal $e_0$ refer to physically different binaries: by the leading-order evolution $e\propto f^{-19/18}$, $e_0=0.3$ at $1\,\mathrm{Hz}$ corresponds to $e\simeq0.05$ at $5\,\mathrm{Hz}$. In the population study $e_0$ is defined at $10\,\mathrm{Hz}$, where the effective expansion parameter at $1\,\mathrm{Hz}$ exceeds unity for $e_0\gtrsim0.09$, so those results extrapolate the truncated expansion and are indicative only. Fourth, the spin-plane study uses the $\mathcal{O}(e_0^2)$ eccentric expansion and the population study the $\mathcal{O}(e_0^4)$ expansion, whereas the single-event results use $\mathcal{O}(e_0^8)$; at the higher eccentricities sampled ($e_0\lesssim0.5$ in the population) the truncated expansions are not converged, so these results should be read as qualitative. Fifth, the template-bank study uses circular-orbit banks constructed independently of the Fisher-forecast parameter ranges; the search-level counterpart of the parameter-estimation gains reported here requires a bank constructed to include eccentricity.

Looking ahead, the framework developed in this work can be extended in several directions. Incorporating spin precession and higher-order harmonics into the eccentric SIQM test would test whether precession-induced phase modulations are degenerate with, or distinguishable from, the eccentricity and quadrupole effects isolated here. Extending the Fisher forecasts to full Bayesian parameter estimation at representative SNRs would quantify how much of the precision reported here survives non-Gaussian noise and realistic priors. Finally, constructing eccentric-spin template banks for aLIGO and 3G detectors would convert the fitting-factor losses of Sec.~\ref{sec:temp_bank} into a direct measure of the detection-efficiency recovered by an eccentric search, closing the gap between the parameter-estimation gains and search-level sensitivity demonstrated here. Since observational bounds on $\kappa_s$ map onto constraints on rotating exotic-compact-object models~\cite{Krishnendu:2025rud}, the measurability demonstrated here suggests that eccentric, spin-sensitive SIQM tests should be a valuable complement to circular-orbit null tests of the binary black hole hypothesis with 3G detectors.

\section{Acknowledgements}
We thank Sumit Kumar for insightful discussions related to template bank generation. We also thank N.V. Krishnendu and K.G. Arun for insightful discussions. K.P also acknowledges the project ``Enabling
Einstein Telescope" with file number OCENW.XL.23.106 of the research programme XL (formerly OC ENW-LARGE) which is (partly) financed by the Dutch Research Council (NWO) under the grant \href{https://doi.org/10.61686/SBOYK19267 }{https://doi.org/10.61686/SBOYK19267}.   
S.N. gratefully acknowledges Poland's high-performance infrastructure PLGrid (HPC Centers: ACK Cyfronet AGH and CI TASK) for providing computer facilities and support within computational grants no.\ PLG/2025/018912 and PLG/2026/019469, respectively. S.N. acknowledges support from the Indian Institute of Technology Madras, where this work was initiated.
S.N. is also thankful to Rimo Das and M. Laxman for useful discussions, and to Unnikrishnan, Subhrata, and Arpita for their encouragement and support.


\section*{Appendix}

\section*{Appendix A: Spin-Plane Analysis at $M=90\,M_\odot$}\label{app:m90}

\begin{figure}[b]
    \centering
    \includegraphics[width=\columnwidth]{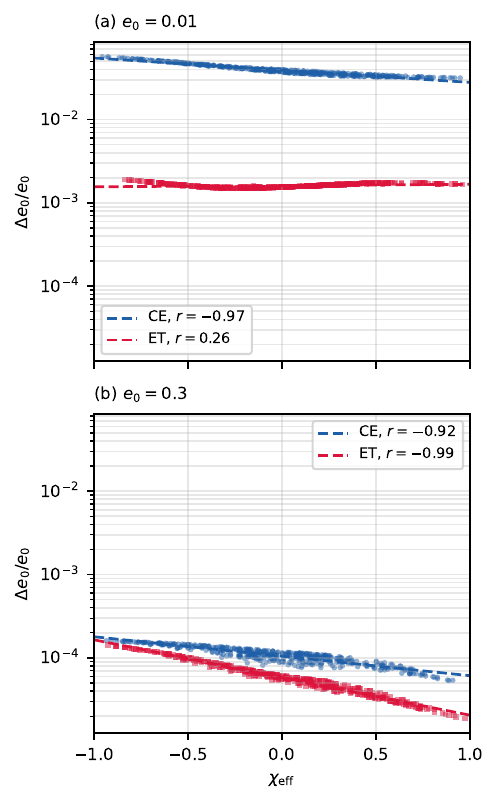}
    \caption{\justifying\textbf{Fractional eccentricity uncertainty versus effective spin at $M=90\,M_\odot$.} As in Fig.~\ref{fig:m10}, for (a)~$e_0=0.01$ and (b)~$e_0=0.3$.}
    \label{fig:m90_appendix}
\end{figure}

Figure~\ref{fig:m90_appendix} and Table~\ref{tab:m90_appendix} give the $M=90\,M_\odot$ counterpart of the $M=10\,M_\odot$ analysis in Sec.~\ref{sec:single_event_results}, with the same $\mathcal{O}(e_0^2)$ truncation, sampling and definitions. The table lists the median errors on $\Delta e_0/e_0$ and $\kappa_s$, their Pearson correlation with $\chi_{\rm eff}$, and the hang-up ratios $\mathcal{R}_{e_0}$ and $\mathcal{R}_{\kappa}$ at $e_0=0.01$ and $0.3$ for CE and ET.

Both detectors improve in $\Delta e_0/e_0$ with eccentricity, CE by a factor of $\simeq350$ ($3.8\times10^{-2}\to1.1\times10^{-4}$) and ET by $\simeq28$ ($1.6\times10^{-3}\to5.7\times10^{-5}$). The CE/ET ratio therefore narrows from $\simeq24$ at $e_0=0.01$ to $\simeq1.9$ at $e_0=0.3$, as at $M=10\,M_\odot$ (where it narrows from $9.7$ to $1.7$), and ET remains the more precise instrument at both eccentricities.

The hang-up ratio $\mathcal{R}_{e_0}$ increases with eccentricity for both detectors (CE: $1.243\to1.382$; ET: $0.933\to1.818$), and the ET correlation changes from essentially flat ($r=0.26$) at $e_0=0.01$ to $r=-0.99$ at $e_0=0.3$. For $\Delta\kappa_s$ the correlation with $\chi_{\rm eff}$ stays weak ($|r|<0.13$ throughout), confirming that the quadratic spin coupling suppresses sensitivity to spin \emph{alignment}. Given this weak correlation, we do not interpret the values of $\mathcal{R}_\kappa$ ($0.86$--$1.35$) as a hang-up effect in $\kappa_s$.

\begin{table}[H]
\caption{\justifying\textbf{Diagnostic summary for $M=90\,M_\odot$.} Median $\Delta e_0/e_0$ and $\Delta\kappa_s$; Pearson coefficients $r_e$ and $r_\kappa$ between $\log_{10}$ of the error and $\chi_{\rm eff}$; hang-up ratios $\mathcal{R}_{e_0} = \Delta e_0(\chi_{\rm eff}<0)/\Delta e_0(\chi_{\rm eff}>0)$ and $\mathcal{R}_{\kappa} = \Delta\kappa_s(\chi_{\rm eff}<0)/\Delta\kappa_s(\chi_{\rm eff}>0)$.}
\label{tab:m90_appendix}
\scriptsize
\setlength{\tabcolsep}{5pt}
\begin{tabular}{l c c c c c c c}
\toprule
Det. & $e_0$ & $\Delta e_0/e_0$ & $r_e$ & $\mathcal{R}_{e_0}$ & $\Delta\kappa_s$ & $r_\kappa$ & $\mathcal{R}_{\kappa}$ \\
\midrule
CE & 0.01 & $3.80\times10^{-2}$ & $-0.971$ & 1.243 & $2.59\times10^{-2}$ & 0.028 & 1.034 \\
CE & 0.30 & $1.08\times10^{-4}$ & $-0.917$ & 1.382 & $2.74\times10^{-2}$ & 0.066 & 0.861 \\
ET & 0.01 & $1.60\times10^{-3}$ & 0.256 & 0.933 & $2.78\times10^{-2}$ & $-0.096$ & 1.082 \\
ET & 0.30 & $5.70\times10^{-5}$ & $-0.991$ & 1.818 & $1.53\times10^{-2}$ & $-0.128$ & 1.354 \\
\bottomrule
\end{tabular}
\end{table}

\clearpage
\bibliographystyle{apsrev}
\bibliography{reference} 
\end{document}